\documentclass[10 pt,slashbox]{article}
\usepackage[letterpaper,hmargin=1in,vmargin=1.5in]{geometry}
\usepackage{indentfirst}
\usepackage{amsmath,amsthm,amssymb}
\usepackage{graphicx}
\usepackage{subfigure}
\usepackage{enumerate}
\usepackage{tikz}

\newtheorem{Theorem}{Theorem}

\newtheorem{Definition}{Definition}
\newtheorem{Example}{Example}
\newtheorem{Remark}{Remark}

\newtheorem{Lemma}{Lemma}

\begin{document}

\title{A Sytematic Piggybacking Design for Minimum Storage Regenerating Codes}
\author{Bin Yang, Xiaohu Tang, \emph{Member,~IEEE}, and  Jie Li
\thanks{The authors are with the Information Security and National Computing Grid Laboratory, Southwest Jiaotong University, Chengdu, 610031, China (e-mail: metroyb@hotmail.com, xhutang@swjtu.edu.cn, jieli873@gmail.com).}}
%\institute{School of Information Science and Technology, Southwest
%Jiaotong University}
\date{}
\maketitle

\textbf{Abstract}--- Piggybacking is an efficient method to decrease the repair bandwidth of  Maximum Distance Separable (MDS) codes or Minimum Storage Regenerating (MSR) codes. In this paper, for minimizing the repair bandwidth of parity nodes of the known MSR codes with high rate, which is usually
the whole size of the original data, i.e., the maximal, a new systematic piggybacking design is proposed
through an in-depth analysis of the design of piggybacking. As a result, new MSR codes are obtained with almost optimal repair bandwidth of parity nodes  while retaining the optimal repair bandwidth  of systematic nodes. Furthermore, MSR codes with balanced download during node repair process are presented based on the new piggybacking design.

\normalfont

\section{Introduction}\label{section_of_Introduction}
Distributed storage systems provide reliable storage service by storing data on distributed storage nodes with redundancy. Since any individual storage node may fail, redundancy is essential to ensure the reliability. Basically, there are two mechanisms of redundancy: replication and erasure coding. Compared with erasure coding, replication is simpler but has lower storage efficiency. Therefore, with data growing much faster than before, erasure coding has been adopted by more and more distributed storage systems, such as
Google Colossus (GFS2) \cite{gfs2}, Microsoft Azure \cite{azure},
HDFS Raid \cite{hdfs-raid}, and OceanStore \cite{oceanstore}.

Maximum Distance Separable (MDS) codes are typical optimal erasure codes in terms of the redundancy-reliability tradeoff. A $(k+r,k)$ MDS storage code, composed of $k+r$ nodes, can tolerate the failure of any $r$ nodes, i.e., any $k$ nodes suffice the reconstruction of the original data. In particular, if the original data is partitioned into $k$ parts and stored in $k$ nodes  without coding, called as systematic nodes, and the other $r$ nodes, termed as parity nodes, store parity data of $k$ nodes, then the $(k+r,k)$ MDS storage code is said to be systematic. In principle, systematic MDS storage codes are preferred in practical systems.

As mentioned above, any individual storage node is not very stable. Once a node fails, we must repair it to maintain the redundancy. Accordingly, the cost to repair a failed storage node is crucial for  evaluating the performance of erasure codes. Generally speaking, there are four metrics
for the cost of node repair, such as computation load, disk I/O, network bandwidth, and the number of accessed disks. The repair bandwidth, defined as the amount of data downloaded to repair a failed node, is the primary concern because the available bandwidth is becoming more and more scarce. Unfortunately, for a $(k+r,k)$ MDS storage code with each node storing $\alpha$ data, the repair bandwidth of a single failed node is equal to $M=k\alpha$, the amount of the whole original data.

Recently, Dimakis \textit{et al.} studied a symmetric repair scenario of $(k+r,k)$  MDS storage codes for distributed storage systems, where
the failed node is repaired by downloading $\beta\le \alpha$ data from each of any $d\ge k$ surviving nodes, i.e., the repair bandwidth is
$\gamma=d\beta$. As a result, they derived a tradeoff between storage and repair bandwidth in \cite{coding}. Codes lying on this tradeoff are called Regenerating Codes, in which MBR  corresponding to minimum repair bandwidth and MSR  corresponding to the minimum storage are the most important.   In this paper, we focus on systematic $(k+r,k)$ MSR codes with high rate, i.e., $r\ll k$. In fact, systematic $(k+r,k)$ MSR codes can be regarded as a special class of systematic $(k+r,k)$ MDS storage codes with minimal repair bandwidth $ \gamma_{\mathrm{MSR}} = d\alpha/(d-k+1)$ where $\beta=\alpha/(d-k+1)$. In order to further reduce the repair bandwidth, we specifically set $d=k+r-1$ throughout this paper, i.e.,
\begin{eqnarray}\label{repair_bandwidth_MSR_bound}
 \gamma_{\mathrm{MSR}} = {k+r-1 \over r}\alpha
\end{eqnarray}
In \cite{InvariantSubspace,hadamard,zigzag,longMDS}, several explicit constructions of high rate MSR codes have been proposed. However, though any of these MSR codes can optimally repair the systematic nodes with respect to the theoretic bound in \eqref{repair_bandwidth_MSR_bound}, all of them except for the $(k+2,k)$ Hadamard MSR code in \cite{hadamard} repair the parity nodes trivially by downloading the whole original data from all the systematic nodes, i.e.,
\begin{eqnarray*}
\gamma_{\mathrm{system}} = {k+r-1 \over r}\alpha,~ \gamma_{\mathrm{parity}} = k\alpha
\end{eqnarray*}

In \cite{piggyback}, the method called piggybacking was presented to reduce the amount of data read and downloaded for node repair of MDS codes and MSR codes. The basic idea of piggybacking is taking multiple instances of a given base code, which can be a MDS code or a MSR code, and adding functions of the data of some instances to the other. Several designs of piggybacking were presented in \cite{piggyback} to improve the repair efficiency of systematic nodes of MDS codes, so were designs of piggybacking for repair of parity nodes of MDS codes and MSR codes as supplement. Consequently, an average saving of  $25\%$ to $50\%$ in the amount of download could be achieved during the node repair. However, this result is still far away from the bound in \eqref{repair_bandwidth_MSR_bound}, which gives $\gamma/M\approx 1/r$ for $r\ll k$.

Inspired by the piggybacking designs in \cite{piggyback} and motivated by the inefficiency of repair of parity nodes of most MSR codes, a systematic analysis on the design of piggybacking in order to minimize the average repair bandwidth of parity nodes of MSR codes is done in this paper. As a result, a new piggybacking design based on given MSR codes which can generate new MSR codes with almost optimal repair bandwidth for parity nodes is proposed. Moreover, since the new piggybacking design doesn't have the property of balanced download during node repair similar to MDS codes, a method is proposed to construct piggybacked codes with balanced download based on the new piggybacking design.

The remainder of this paper is organized as follows. In Section \ref{section_of_MSR}, the model of systematic MSR codes and the piggybacking design for repair of parity nodes of MSR codes presented in \cite{piggyback} are briefly reviewed. In Section \ref{section_of_construction}, a detailed analysis of the design of piggybacking is elaborated and consequently our new piggybacking design is proposed. Based on the new piggybacking design, a method to construct piggybacked codes with balanced download is proposed in Section \ref{section_symmetric_piggyback}. Finally, the conclusion
is given in Section \ref{section_of_conclusion}.

\section{Systematic $(k+r,k)$  MSR code and Piggybacking}\label{section_of_MSR}

A systematic $(k+r,k)$ MSR code consists of $k$ systematic nodes and $r$ parity nodes.  Assume that the amount of the original data is $M=k\alpha'$, it can be equally partitioned into $k$ parts $\textbf{f}=[\mathbf{f}_1^T,\mathbf{f}_2^T,\cdots,\mathbf{f}_k^T]^T$ and placed on $k$ systematic nodes, where $\mathbf{f}_i$ is an $\alpha'\times 1$ vector.  In general, $r$ parity nodes hold parity data, namely $r$ vectors $\mathbf{f}_{k+1},\cdots,\mathbf{f}_{k+2}$,  of all the systematic data $\mathbf{f}_{1},\cdots,\mathbf{f}_{k}$. Precisely, for $1\le j\le r$, the $j$th parity data is a linear combination of all the systematic data $\mathbf{f}_{1},\cdots,\mathbf{f}_{k}$
as
\begin{eqnarray*}%\label{MSR_Code_Def}
\mathbf{f}_{k+j}=A_{j,1}\mathbf{f}_1+\cdots + A_{j,k}\mathbf{f}_k
\end{eqnarray*}
where the matrix $A_{j,i}$, $1\le i\le k$ of order $\alpha'\times\alpha'$, is called the \emph{coding matrix} of the $j$th parity node for the $i$th systematic node.  Table \ref{Hadamard_Model} illustrates the structure of such a $(k+r,k)$ MSR code.
\begin{table}[htbp]
\begin{center}
\caption{Structure of a $(k+r,k)$ MSR code}\label{Hadamard_Model}
\begin{tabular}{|c|c|}
\hline
Systematic node & Systematic data \\
\hline
1 & $\mathbf{f}_1$ \\
\hline
\vdots & \vdots \\
\hline
$k$ & $\mathbf{f}_k$ \\
\hline
Parity node & Parity data \\
\hline
$1$ & $\mathbf{f}_{k+1}=A_{1,1}\mathbf{f}_1+ \cdots+ A_{1,k}\mathbf{f}_k$ \\
\hline
\vdots & \vdots \\
\hline
$r$ & $\mathbf{f}_{k+r}=A_{r,1}\mathbf{f}_1+\cdots + A_{r,k}\mathbf{f}_k$ \\
\hline
\end{tabular}
\end{center}
\end{table}

Recall that all the known MSR codes have the optimal repair ability to repair the systematic
nodes.  Once the $i$th systematic node fails, one downloads  data $S_{i,j}\mathbf{f}_j$, $1\le j\neq i\le k+r$, from all the surviving
nodes by an $\frac{\alpha'}{r}\times\alpha'$ matrix  $S_{i,j}$ of rank $\frac{\alpha'}{r}$, and then recover
the original data $\mathbf{f}_i$. That is, only a
proportion $1/r$ of data
is needed from each of other $k+r-1$ nodes to repair a failed  systematic node.  So, totally $(k+r-1)\alpha'/r$ data is downloaded, which is optimal with
respect to the theoretic bound in \eqref{repair_bandwidth_MSR_bound}. But
for repair of a failed parity node, all the known MSR codes have to download all the $\alpha'$ data from each of $k$ systematic nodes, i.e., totally $M$ data, much bigger than the optimal value, except for $(k+2,k)$ Hadamard MSR code whose parity nodes can be repaired similar to the systematic nodes.

In the rest of this paper, we always assume $S_{i,j}=S_i$ for all $1\le i\le k$, $1\le j\ne i\le k+r$ to simplify the repair strategy, and call it the \emph{repair matrix} of the $i$th systematic node. It should be noted that all the known constructions have such repair matrices, for example Zigzag code \cite{zigzag}, Hadamard MSR code \cite{hadamard}, long MDS code \cite{longMDS}, invariant subspace codes \cite{InvariantSubspace}, etc.

In \cite{piggyback}, a piggybacking design devoted to efficient repair of parity nodes of MSR codes was presented. By taking
two instances of a systematic $(k+r,k)$  MSR code and denoting by $\mathbf{f}_j^{(i)}$ the data of $j$th node in instance $i$, $1\le j\le k+r$ and $1\le i\le 2$, the piggybacking design is illustrated in Table \ref{Tab_pigg} where the only piggyback is deployed on the first node of instance $2$, which is the sum of all parity data of instance $1$  except the first one.

\begin{table}
\begin{center}
\caption{Structure of a piggybacking for $(k+r,k)$ MSR code}\label{Tab_pigg}
\begin{tabular}{|c|c|c|}
\hline
                 & Instance 1 & Instance 2 \\
\hline
Systematic Node & Systematic data & Systematic data\\
\hline
 1                 & $\mathbf{f}_1^{(1)}$ & $\mathbf{f}_1^{(2)}$ \\
\hline
$\vdots$  & $\vdots$ & $\vdots$ \\
\hline
$k$                 & $\mathbf{f}_k^{(1)}$ & $\mathbf{f}_k^{(2)}$ \\
\hline
Parity Node & Parity data & Parity data\\
\hline
1 & $\mathbf{f}_{k+1}^{(1)}$ & $\mathbf{f}_{k+1}^{(2)}+\sum_{j=2}^r{\mathbf{f}_{k+j}^{(1)}}$ \\
\hline
$2$              & $\mathbf{f}_{k+2}^{(1)}$ & $\mathbf{f}_{k+2}^{(2)}$ \\
\hline
  $\vdots$        & \vdots & \vdots \\
\hline
$r$                 & $\mathbf{f}_{k+r}^{(1)}$ & $\mathbf{f}_{k+r}^{(2)}$ \\
\hline
\end{tabular}
\end{center}
\end{table}

The resultant piggybacked $(k+r,k)$ MSR code stores $\alpha=2\alpha'$ data at each node.
In principle, there are two distinct repair strategies for parity nodes:
\begin{enumerate}
\item [(1)] To repair the first parity node, one downloads all the systematic data of the two instances, totally $2k\alpha'$ data;
\item [(2)] To repair the $i$th parity node, $i\ne 1$, one downloads all the systematic data of  instance 2, the parity data
$\mathbf{f}_{k+2}^{(1)},\cdots,\mathbf{f}_{k+i-1}^{(1)},\mathbf{f}_{k+i+1}^{(1)},\cdots,\mathbf{f}_{k+r}^{(1)}$ of instance 1,
and the data $\mathbf{f}_{k+1}^{(2)}+\sum_{j=2}^r{\mathbf{f}_{k+j}^{(1)}}$,
totally $(k+r-1)\alpha'$ data.
\end{enumerate}
Hence, the average repair bandwidth of the parity nodes is
\begin{eqnarray}\label{old_parity_design}
\gamma_{\textrm{parity}} &=& {2k\alpha'+(r-1)(k+r-1)\alpha'\over r}={2k\alpha+(r-1)(k+r-1)\alpha\over 2r}
\end{eqnarray}
which is
far away from the bound in  \eqref{repair_bandwidth_MSR_bound}, especially when $k\gg r$, however.

Throughout this paper, we always assume that $k\gg r$ since  the high rate MSR codes with $k\gg r$ are usually of great interest. In this case, the $2k\alpha'$ data downloaded in the first strategy is much larger than $(k+r-1)\alpha'$ data downloaded in the second strategy. Note from
Table \ref{Tab_pigg} that the larger downloading is caused by the missing of $\mathbf{f}_{k+1}^{(1)}$ in the piggyback so that one has to download
all the $k\alpha'$ systematic data of instance 1 instead of using less (at most $2(r-1)\alpha'$) parity data in the $r-1$ parity nodes and the piggyback. This observation
immediately gives us a hint that $\mathbf{f}_{k+1}^{(1)}$ should be included in a piggyback. As an example shown in Table \ref{Tab_exam},   the average repair bandwidth of the parity nodes can be reduced as
\begin{eqnarray*}
\gamma_{\textrm{parity}}' &=& {2(k+r)\alpha'+(r-2)(k+r-1)\alpha'\over r}={(k+r+1)\alpha'+(r-1)(k+r-1)\alpha'\over r}< \gamma_{\textrm{parity}}
\end{eqnarray*}
Inspired by the effect of the new piggybacking, we will discuss how to design piggybacking systematically for minimizing the average  repair bandwidth of the parity nodes in the next section.

\begin{table}
\begin{center}
\caption{Structure of a modified piggybacking for $(k+r,k)$ MSR code}\label{Tab_exam}
\begin{tabular}{|c|c|c|}
\hline
                 & Instance 1 & Instance 2 \\
\hline
Systematic Node & Systematic data & Systematic data\\
\hline
 1                 & $\mathbf{f}_1^{(1)}$ & $\mathbf{f}_1^{(2)}$ \\
\hline
$\vdots$  & $\vdots$ & $\vdots$ \\
\hline
$k$                 & $\mathbf{f}_k^{(1)}$ & $\mathbf{f}_k^{(2)}$ \\
\hline
Parity Node & Parity data & Parity data\\
\hline
1 & $\mathbf{f}_{k+1}^{(1)}$ & $\mathbf{f}_{k+1}^{(2)}+\sum_{j=2}^r{\mathbf{f}_{k+j}^{(1)}}$ \\
\hline
$2$              & $\mathbf{f}_{k+2}^{(1)}$ & $\mathbf{f}_{k+2}^{(2)}+\mathbf{f}_{k+1}^{(1)}$ \\
\hline
  $\vdots$        & \vdots & \vdots \\
\hline
$r$                 & $\mathbf{f}_{k+r}^{(1)}$ & $\mathbf{f}_{k+r}^{(2)}$ \\
\hline
\end{tabular}
\end{center}
\end{table}

\begin{Remark}
In \cite{piggyback}, a smaller average repair bandwidth, but still much larger than  the bound in  \eqref{repair_bandwidth_MSR_bound}, can be obtained by  partitioning the $r$ parity nodes into $g=\max(1,\lfloor{r\over {\sqrt{k+1}}}\rfloor)$ groups as equally as possible, i.e., each of group $1$ to group $g-1$ has $h=\lfloor{r\over g}\rfloor$ nodes and group $g$ has $h'=r-(g-1)h$ nodes. For simplicity we only consider $g=1$ herein due to
$k\gg r$.
\end{Remark}

\section{New Piggybacking for Parity Repair of $(k+r,k)$ MSR code}\label{section_of_construction}

In this section, we present a general transform for MSR codes based on the piggybacking method, which can give
MSR codes with almost optimal repair bandwidth of parity nodes.

Consider $s$ instances of a $(k+r,k)$ MSR code, where $2\le s\le r$. Denote by vectors $\mathbf{f}_{1}^{(i)}, \cdots, \mathbf{f}_{k}^{(i)}$ and  $\mathbf{f}_{k+1}^{(i)}, \cdots, \mathbf{f}_{k+r}^{(i)}$ respectively the systematic data and parity data of the $i$th instance of the MSR code. Next, we apply the piggybacking method to the $s$ instances:
\begin{itemize}
\item Keep the first $s-1$ instances unchanged;
\item Keep the systematic data of the instance $s$ unchanged but add to the $i$-th parity data $\mathbf{f}_{k+i}^{(s)}$, $1\le i\le r$, the piggyback $\mathbf{P}_i$, which is a linear combination
of  $\mathbf{f}_{k+l}^{(j)}$ for $1\le l\ne i\le r$ and $1\le j<s$, i.e.,
\begin{eqnarray}\label{Eq_P1-r}
(\mathbf{P}_1, \cdots, \mathbf{P}_r)
= (\mathbf{f}_{k+1}^{(1)}, \cdots,
\mathbf{f}_{k+1}^{(s-1)})A_1+\cdots+
(\mathbf{f}_{k+r}^{(1)}, \cdots, \mathbf{f}_{k+r}^{(s-1)})A_r
\end{eqnarray}
where $A_l$ is a $(s-1)\times r$ matrix of rank $s-1$ with the $l$th column being the all-zero column for any $1\le l\le r$. For convenience, we call $A_1,\cdots,A_r$ piggybacking matrices which define the piggyback set $\{\mathbf{P}_1,\cdots,\mathbf{P}_{r}\}$.
\end{itemize}
Let $\alpha'$ denotes the data amount of a node of the original $(k+r,k)$ MSR code. Then, we get a  $s$-piggybacked $(k+r,k)$ MSR code
having $k$ systematic nodes and $r$ parity nodes, each storing $\alpha=s\alpha'$ data, whose structure is depicted in Table \ref{Instance-s}.

\newcommand{\rb}[1]{\raisebox{1.5ex}[0pt]{#1}}
\begin{table}[htbp]
\begin{center}
\caption{Structure of a piggybacking for $(k+r,k)$ MSR code}\label{Instance-s}
\begin{tabular}{|c|c|c|c|c|}
\hline
& Instance $1$ &  Instance $2$ & $\cdots$ &  Instance $s$\\
\hline
Systematic &Systematic & Systematic  &  & Systematic \\
node & data & data & \rb{$\cdots$} & data\\
\hline
1 & $\mathbf{f}_1^{(1)}$ & $\mathbf{f}_1^{(2)}$ &$\cdots$ &$\mathbf{f}_1^{(s)}$\\
\hline
$\vdots$ & $\vdots$ &$\vdots$ & $\ddots$& $\vdots$ \\
\hline
$k$ & $\mathbf{f}_k^{(1)}$& $\mathbf{f}_k^{(2)}$ &$\cdots$ &$\mathbf{f}_k^{(s)}$ \\
\hline
Parity & Parity & Parity  & & Parity\\
node & data & data & \rb{$\cdots$} & data\\
\hline
$1$ & $\mathbf{f}_{k+1}^{(1)}$ & $\mathbf{f}_{k+1}^{(2)}$ &$\cdots$ &$\mathbf{f}_{k+1}^{(s)}+\mathbf{P}_1$\\
\hline
$\vdots$ & $\vdots$ &$\vdots$ & $\ddots$& $\vdots$ \\
\hline
$r$ & $\mathbf{f}_{k+r}^{(1)}$& $\mathbf{f}_{k+r}^{(2)}$ &$\cdots$ &$\mathbf{f}_{k+r}^{(s)}+\mathbf{P}_{r}$ \\
\hline
\end{tabular}
\end{center}
\end{table}

\begin{Theorem}\label{theorem_piggyback_MDS}
The $s$-piggybacked $(k+r,k)$ MSR code has MDS property.
\end{Theorem}
\textit{Proof}: For any $k$ nodes out of all the $k+r$ nodes,
if they are all systematic nodes, then we are done. Otherwise, noting that the first $s-1$ instances
are unchanged, we can reconstruct all their  systematic data by
means of the MDS property of the original MSR code. Next, we cancel the parity data of the first $s-1$ instances
in the involved piggybacks with the help of the systematic data of the first $s-1$ instances so that we are able
to  reconstruct  the unchanged systematic data of the last instance still  by the MDS property of the original MSR code.

\hfill$\Box$

Specifically, if the piggyback set $\{\mathbf{P}_1,\cdots,\mathbf{P}_{r}\}$ leads to the minimal average repair bandwidth of the parity nodes
of a $s$-piggybacked $(k+r,k)$ MSR code, then it is said to be \textit{optimal}.
From now on, we focus on the design of the optimal piggyback set $\{\mathbf{P}_1,\cdots,\mathbf{P}_{r}\}$. For a $s$-piggybacked $(k+r,k)$ MSR code, recall from the discussion
in the previous section that we already have

\vspace*{.1cm}
\setlength{\parindent}{0pt}
\textbf{Principle of Repair of Parity Nodes}: If parity node $1\le i\le r$ fails, then the parity data
$\mathbf{f}_{k+i}^{(1)},\cdots,\mathbf{f}_{k+i}^{(s-1)}$ are repaired as follows:
\begin{enumerate}
\item [Step 1] Download the systematic data $\mathbf{f}_{1}^{(s)},\cdots,\mathbf{f}_{k}^{(s)}$ of instance $s$, and calculate $\mathbf{f}_{k+1}^{(s)},\cdots,\mathbf{f}_{k+r}^{(s)}$;
\item [Step 2] Recover each of the parity data $\mathbf{f}_{k+i}^{(1)},\cdots,\mathbf{f}_{k+i}^{(s-1)}$ from \eqref{Eq_P1-r} by downloading involved parity data $\mathbf{f}_{k+j}^{(s)}+\mathbf{P}_j$ from instance $s$ and other parity data involved in $\mathbf{P}_j$ from instances $1,\cdots,s-1$, with $\mathbf{f}_{k+j}^{(s)}$ calculated in Step 1, for some integer $1\le j\ne i\le r$;
\item [Step 3] Recover $\mathbf{f}_{k+i}^{(s)}+\mathbf{P}_i$  from \eqref{Eq_P1-r} by downloading all parity data involved in $\mathbf{P}_i$ from instances $1,\cdots,s-1$, with $\mathbf{f}_{k+i}^{(s)}$ calculated in Step 1.
\end{enumerate}

\vspace*{.1cm}
\setlength{\parindent}{15pt}
If a piggyback $\mathbf{p}$ is composed of parity data including $\mathbf{f}$, then we say $\mathbf{f}$ appears in $\mathbf{p}$. Now, we prove

\begin{Lemma}\label{lemma_optimal_piggyback_matrix}
There exists an optimal piggyback set $\{\mathbf{P}_1,\cdots,\mathbf{P}_{r}\}$ such that any parity data $\mathbf{f}_{k+i}^{(j)}$ for $1\le j<s$ and $1\le i\le r$ appears exactly once. More precisely, the piggybacking matrix $A_i$ in \eqref{Eq_P1-r} is formed by an identity matrix of order $s-1$ and $r-s+1$ zero columns for all $1\le i\le r$.
\end{Lemma}

\textit{Proof}: Suppose that parity node 1 fails. By the Principle of Repair of Parity Nodes,
$\textrm{rank}(A_1)=s-1$ is necessary to  recover $\mathbf{f}_{k+1}^{(1)},\cdots,\mathbf{f}_{k+1}^{(s-1)}$ from \eqref{Eq_P1-r}.  That is, there are $s-1$ independent columns in $A_1$. Denote by $l_1,\cdots,l_{s-1}$ the indices of these  independent columns respectively, and
let $\{1,\cdots,r\}\setminus\{l_1,\cdots,l_{s-1}\}=\{l_{s},\cdots,l_{r}\}$ . Then, after removing $\mathbf{f}_{k+1}^{(1)},\cdots,\mathbf{f}_{k+1}^{(s-1)}$ from the piggybacks $\mathbf{P}_{l_{s}},\cdots,\mathbf{P}_{l_{r}}$, we get new piggybacks as
\begin{eqnarray*}
(\mathbf{P}_1', \cdots, \mathbf{P}_r')
= (\mathbf{f}_{k+1}^{(1)}, \cdots,
\mathbf{f}_{k+1}^{(s-1)})A_1'+(\mathbf{f}_{k+2}^{(1)}, \cdots,
\mathbf{f}_{k+2}^{(s-1)})A_2+\cdots+
(\mathbf{f}_{k+r}^{(1)}, \cdots, \mathbf{f}_{k+r}^{(s-1)})A_r
\end{eqnarray*}
where $A_1'$ is formed from $A_1$ by replacing its columns $l_{s},\cdots,l_{r}$ with
all-zero columns. Obviously, the
new piggybacks would not increase the repair bandwidth  since the repair process of any parity node $1\le i\le r$
does not incur more download to cancel the interference data from other parity nodes in the new piggybacks.

Further,  we modify piggybacks  as
\begin{eqnarray*}
(\mathbf{P}_{1}'', \cdots, \mathbf{P}_{r}'')
= (\mathbf{f}_{k+1}^{(1)}, \cdots,
\mathbf{f}_{k+1}^{(s-1)})A_1''+(\mathbf{f}_{k+2}^{(1)}, \cdots,
\mathbf{f}_{k+2}^{(s-1)})A_2+\cdots+
(\mathbf{f}_{k+r}^{(1)}, \cdots, \mathbf{f}_{k+r}^{(s-1)})A_r
\end{eqnarray*}
where  $A_1''$ is formed by replacing the submatrix of $A_1'$ consisting of its columns $l_{1},\cdots,l_{s-1}$
with the identity matrix. It is easily checked that in contrast to   $\{\mathbf{P}_{1}', \cdots, \mathbf{P}_{r}'\}$,
the new piggyback set $\{\mathbf{P}_{1}'', \cdots, \mathbf{P}_{r}''\}$ keeps the same repair bandwidth
for the repair of parity node $1$, and does not increase the repair bandwidth for the repair of parity node $2\le i\le r$
since no more download is needed to cancel the interference data from other parity nodes.

Recursively applying the above procedure to node 2 to $r$, we can change the original piggybacks to
those having  the piggybacking matrices in the desired form, whose average repair bandwidth of parity nodes is no more than that of the former.
This completes the proof.

\hfill$\Box$

Regarding the optimal piggyback set in Lemma \ref{lemma_optimal_piggyback_matrix},
denote $p_i(j)=l$ if $\mathbf{f}_{k+i}^{(j)}$,  $1\le i\le r, 1\le j<s$,
appears in the piggyback $\mathbf{P}_l$, $1\le l\le r$.
In particular, set  $p_i(s)=i$.
Then, $p_i$ is an injective function from $\{1,\cdots,s\}$ into $\{1,\cdots,r\}$ for all $1\le i\le r$.
Accordingly, we can rewrite the piggyback as
\begin{eqnarray}\label{Eqn_P_SF}
\mathbf{P}_i=\sum_{(i',j')\in S_i} \mathbf{f}_{k+i'}^{(j')}
\end{eqnarray}
where $S_i=\{(i',j')| p_{i'}(j')=i, 1\le i'\le r, 1\le j'<s\}$.

Define
$L(\mathbf{P}_i)=|S_i|$. We have the following lemma about  $L(\mathbf{P}_i)$.

\begin{Lemma}\label{lemma_optimal_piggyback_length}
A piggyback set $\{\mathbf{P}_1,\cdots,\mathbf{P}_{r}\}$ defined in \eqref{Eqn_P_SF} is optimal if and only if
$L(\mathbf{P}_{i})=s-1$, $1\le i\le r$.
\end{Lemma}
\textit{Proof}: Suppose that the parity node $1\le i\le r$ fails. According to the Principle of Repair of Parity Nodes, one has to download
\begin{itemize}
\item [Step 1] $k\alpha$ systematic data from instance $s$ and then calculate $\mathbf{f}_{k+1}^{(s)},\cdots,\mathbf{f}_{k+r}^{(s)}$;
\item [Step 2]  $\mathbf{P}_{l}+\mathbf{f}_{k+l}^{(s)}$ and all the data $\mathbf{f}_{k+i'}^{(j')}$ where $(i',j')\in S_l\setminus\{(i,j)\}$ and $p_i(j)=l$ to repair
$\mathbf{f}_{k+i}^{(j)}$, $1\le j<s$, from \eqref{Eqn_P_SF};
\item [Step 3]  All the data $\mathbf{f}_{k+i'}^{(j')}$ where $(i',j')\in S_i$ to repair
$\mathbf{P}_i+\mathbf{f}_{k+i}^{(s)}$ from \eqref{Eqn_P_SF}.
\end{itemize}

It follows from Lemma  \ref{lemma_optimal_piggyback_matrix} that $S_{i_1}\cap S_{i_2}=\varnothing$ if $1\le i_1\ne i_2\le r$. So, the repair
bandwidth of parity node $i$ is
\begin{eqnarray*}
\gamma=k\alpha'+\sum_{j=1}^{s} L(\mathbf{P}_{p_i(j)})\alpha'
\end{eqnarray*}
and then the average repair bandwidth of the parity nodes is
\begin{eqnarray*}
\gamma_{\textrm{parity}}=k\alpha'+{\sum_{i=1}^r \sum_{j=1}^{s} L(\mathbf{P}_{p_i(j)})\over r}\alpha'
\end{eqnarray*}

Next, we calculate $\sum_{i=1}^r \sum_{j=1}^{s} L(\mathbf{P}_{p_i(j)})$. In fact, we
can see that the term $L(\mathbf{P}_{i})$ appears exactly $L(\mathbf{P}_{i})+1$ times in this sum
since the repair of $\mathbf{P}_{i}+\mathbf{f}_{k+i}^{(s)}$ or $\mathbf{f}_{i'}^{(j')}$ needs to download
$L(\mathbf{P}_{i})$ data according to Steps 2 and 3 where $1\le i'\le r$ and $1\le j'<s$ satisfy
$p_{i'}(j')=i$. Thus,
\begin{eqnarray*}
\sum_{i=1}^r \sum_{j=1}^{s} L(\mathbf{P}_{p_i(j)})=\sum_{i=1}^r L(\mathbf{P}_{i})^2+\sum_{i=1}^r L(\mathbf{P}_{i})
\end{eqnarray*}

When the piggyback set is optimal, its $\gamma_{\textrm{parity}}$ must be minimized, so
does $\sum_{i=1}^r L(\mathbf{P}_{i})^2+\sum_{i=1}^r L(\mathbf{P}_{i})$. Note from  Lemma \ref{lemma_optimal_piggyback_matrix}
that $\sum_{i=1}^r L(\mathbf{P}_{i})=r(s-1)$. Then, the piggyback set is optimal
if and only if $L(\mathbf{P}_{i})=s-1$ for all $1\le i\le r$ by the well-known Cauchy-Schwarz inequality.

\hfill$\Box$

Based on Lemmas \ref{lemma_optimal_piggyback_matrix} and \ref{lemma_optimal_piggyback_length},
we are ready to construct a class of optimal piggyback sets for $s$-piggybacked $(k+r,k)$ MSR codes, i.e.,
the piggyback sets $\{\mathbf{P}_1,\cdots,\mathbf{P}_{r}\}$ given in \eqref{Eqn_P_SF} with
the constraint that $L(\mathbf{P}_{i})=s-1$ for all $1\le i\le r$.
In what follows, we give the concrete repair strategy for node failure of the $s$-piggybacked $(k+r,k)$ MSR code.
Precisely, we use two distinct repair strategies to respectively deal with the failure of a systematic node and the failure of a parity node.

(I) When systematic node $i$ fails, $1\le i\le k$,  we repair it by the following steps:
\begin{enumerate}[1)]
    \item [Step 1] Download data $S_{i}\mathbf{f}_{l}^{(j)}$ from nodes $1\le l\ne i\le k$ of each instance $1\le j\le s$, data $S_{i}\mathbf{f}_{k+l}^{(j)}$ from parity nodes $1\le l\le r$ of each instance $1\le j<s$, and data $S_{i}(\mathbf{f}_{k+l}^{(s)}+\mathbf{P}_l)$ from parity nodes $1\le l\le r$ of instance $s$, by the repair matrix $S_{i}$;

    \item [Step 2] Compute $S_{i}\mathbf{P}_{l}$ according to \eqref{Eqn_P_SF}, and then get $S_{i}\mathbf{f}_{k+l}^{(s)}$ by cancelling the piggyback term $S_{i}\mathbf{P}_{l}$ from $S_{i}(\mathbf{f}_{k+l}^{(s)}+\mathbf{P}_l)$, for $1\le l\le r$.

    \item [Step 3] Recover data $\mathbf{f}_{i}^{(j)}$ from all the data $S_{i}\mathbf{f}_{l}^{(j)}$, $1\le l\ne i\le k+r$, by using the repair method of the original MSR code, for each instance $1\le j\le s$.
\end{enumerate}

  \vspace{1mm}

 (II) When parity node $i$ fails, $1\le i\le r$,  we repair it as follows.

 \begin{enumerate}[1)]
 \item [Step 1] Download all the systematic data  $\mathbf{f}_{1}^{(s)},\cdots, \mathbf{f}_{k}^{(s)}$ of instance $s$ and compute $\mathbf{f}_{k+1}^{(s)},\cdots, \mathbf{f}_{k+r}^{(s)}$;

 \item [Step 2] Download $\mathbf{f}_{k+p_i(j)}^{(s)}+\mathbf{P}_{p_i(j)}$ and the other $s-2$ parity data $\mathbf{f}_{k+i'}^{(j')}$ where $(i',j')\in S_{p_i(j)}\setminus\{i,j\}$ to recover the failed data $\mathbf{f}_{k+i}^{(j)}$ for each $1\le j<s$;
 \item [Step 3] Download all $\mathbf{f}_{k+i'}^{(j')}$ where $(i',j')\in S_i$  to recover $\mathbf{f}_{k+i}^{(s)}+\mathbf{P}_i$;
\end{enumerate}

Then, the $s$-piggybacked $(k+r,k)$ MSR code has two kinds of repair bandwidth
\begin{eqnarray*}
\gamma_{\textrm{system}} &=& {(k+r-1)\alpha\over r}
\end{eqnarray*}
and
\begin{eqnarray*}
\gamma_{\textrm{parity}}=(k+s(s-1))\alpha'={(k+s(s-1))\alpha\over s}
\end{eqnarray*}
Obviously, the former is optimal and the latter is not with respect to the bound in \eqref{repair_bandwidth_MSR_bound}. But compared with the original
$(k+r,k)$ MSR code, the $s$-piggybacked $(k+r,k)$ MSR code maintains optimal repair property of the systematic nodes,
and reduces the repair bandwidth of the parity nodes dramatically.
Most notably, the  repair bandwidth of the parity nodes in the $r$-piggybacked $(k+r,k)$ MSR approaches the optimal value in  \eqref{repair_bandwidth_MSR_bound} when
$k$ tends to infinity.

\begin{Theorem}\label{theorem_piggyback_repair}
The $r$-piggybacked $(k+r,k)$ MSR code constructed by $r$ instances has almost optimal repair property:
\begin{enumerate}
\item [(1)] Any systematic node has optimal repair property with  repair  bandwidth $(k+r-1)\alpha/r$;
\item [(2)] Any parity node has almost optimal repair property with repair bandwidth $(k+r(r-1))\alpha/r$
\end{enumerate}
\end{Theorem}
%\textit{Proof}:
%   From the repair strategy of systematic nodes and the proof of Lemma \ref{lemma_optimal_piggyback_length}, we see that $\gamma_{sys}=s(k+r-1)\alpha/r=s(k+r-1)\alpha'/r^2$
%   and $\gamma_{par}=(k/s+s-1)\alpha'$ are needed to repair a failed systematic node and a failed parity node respectively, which turn into $\gamma_{sys}=(k+r-1)\alpha'/r$ and $\gamma_{par}=(k+r(r-1))\alpha'/r$ when $s=r$. According to the theoretic bound in \eqref{repair_bandwidth_MSR_bound}, the optimal repair bandwidth should be $(k+r-1)\alpha'/r$, so the assertion is proved.
%
%\hfill$\Box$

From the above analysis,  the injections $p_i$, $1\le i\le r$, which result in $|S_l|=s-1$ for
all $1\le l\le r$, are crucial to the optimal piggyback set. Fortunately, there are plenty of
such functions. For example,
\begin{eqnarray}\label{Eqn_MainD}
p_i(j)=((i-j+s-1)~\bmod~r)+1
\end{eqnarray}
and
\begin{eqnarray}\label{Eqn_MinorD}
p_i(j)=((i+j-s-1)~\bmod~r)+1
\end{eqnarray}
are  two classes of injective functions from $\{1,\cdots,s\}$ into $\{1,\cdots,r\}$ obtained from the diagonals of slopes 1 and -1 of a $r\times s$ matrix.

Finally, we demonstrate two illustrative examples of the optimal piggyback set for $(k+4,k)$ MSR codes.

\begin{Example} Two optimal piggyback sets  for $3$-piggybaced $(k+4,k)$ MSR codes are respectively
\begin{eqnarray*}
\mathbf{P}_1=\mathbf{f}_{k+3}^{(1)}+\mathbf{f}_{k+4}^{(2)},~\mathbf{P}_2=\mathbf{f}_{k+4}^{(1)}+\mathbf{f}_{k+1}^{(2)},
~\mathbf{P}_3=\mathbf{f}_{k+1}^{(1)}+\mathbf{f}_{k+2}^{(2)},~\mathbf{P}_4=\mathbf{f}_{k+2}^{(1)}+\mathbf{f}_{k+3}^{(2)}
\end{eqnarray*}
and
\begin{eqnarray*}
\mathbf{P}_1=\mathbf{f}_{k+3}^{(1)}+\mathbf{f}_{k+2}^{(2)},~\mathbf{P}_2=\mathbf{f}_{k+4}^{(1)}+\mathbf{f}_{k+3}^{(2)},
~\mathbf{P}_3=\mathbf{f}_{k+1}^{(1)}+\mathbf{f}_{k+4}^{(2)},~\mathbf{P}_4=\mathbf{f}_{k+2}^{(1)}+\mathbf{f}_{k+1}^{(2)}
\end{eqnarray*}
which are based on the injective functions \eqref{Eqn_MainD} and \eqref{Eqn_MinorD} respectively. In fact, these two injective functions can be
  obtained from the diagonals of slopes 1 and -1 of
a $4\times 3$ matrix as depicted in Figure 1.

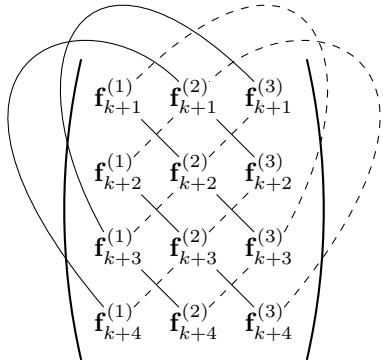
\begin{figure}[!hbp]
\begin{center}
\begin{tikzpicture}
\clip (-1,0) rectangle (4,4.8);
%brakets
\draw [thick] (0,0) .. controls (-0.3,1.3) and (-0.3,2.7) .. (0,4);
\draw [thick] (3,0) .. controls (3.3,1.3) and (3.3,2.7) .. (3,4);
\foreach \i in {1, 2, 3, 4}
{\foreach \j in {1, 2, 3}
    \draw (\j - 1 + 0.5, 4 - \i + 0.5) node{$\mathbf{f}_{k+\i}^{(\j)}$};
}
%for slope = 1
\foreach \i in {2, 3}
{
    \draw (1 - 0.2, \i + 0.2) -- (1 + 0.3, \i - 0.3);
    \draw (2 - 0.2, \i - 1 + 0.2) -- (2 + 0.3, \i - 1 - 0.3);
}
\draw (2 - 0.2, 3 + 0.2) -- (2 + 0.3, 3 - 0.3);
\draw (1 - 0.2, 1 + 0.2) -- (1 + 0.3, 1 - 0.3);
\draw (0 + 0.3, 2 - 0.3) .. controls (-1,4) and (0,6) .. (2 + 0.3, 4 - 0.3);
\draw (0 + 0.3, 1 - 0.3) .. controls (-2.5,4) and (0,5) .. (1 + 0.3, 4 - 0.3);
%for slope = -1
\foreach \i in {0, 1}
{
    \draw [dashed] (\i + 0.7, \i + 0.7) -- (\i + 1 + 0.3, \i + 1 + 0.3);
    \draw [dashed] (\i + 0.7, \i + 1 + 0.7) -- (\i + 1 + 0.3, \i + 2 + 0.3);
}
\draw [dashed] (0 + 0.7, 2 + 0.7) -- (1 + 0.3, 3 + 0.3);
\draw [dashed] (1 + 0.7, 0 + 0.7) -- (2 + 0.3, 1 + 0.3);
\draw [dashed] (2 + 0.7, 1 + 0.7) .. controls (4,4) and (3,6) .. (0 + 0.7, 3 + 0.7);
\draw [dashed] (2 + 0.7, 0 + 0.7) .. controls (5.5,4) and (3,5) .. (1 + 0.7, 3 + 0.7);
\end{tikzpicture}
\end{center}
\caption{Two injective functions of a $3$-piggybacked $(k+4,k)$ MSR code}
\end{figure}

By the two  optimal piggyback sets, the repair bandwidth of the parity nodes is
$(k+6)\alpha/3$ where $\alpha=3\alpha'$. Recall that the optimal repair bandwidth of $(k+4,k)$ MSR codes given in \eqref{repair_bandwidth_MSR_bound} is $(k+3)\alpha/4$.
Further, we can approach the optimal value by adding another instance to get $4$-piggybaced $(k+4,k)$ MSR codes.
3 optimal piggyback sets are listed for $4$-piggybaced $(k+4,k)$ MSR codes.
\begin{eqnarray*}
\mathbf{P}_1=\mathbf{f}_{k+2}^{(1)}+\mathbf{f}_{k+3}^{(2)}+\mathbf{f}_{k+4}^{(3)},~\mathbf{P}_2=\mathbf{f}_{k+3}^{(1)}
+\mathbf{f}_{k+4}^{(2)}+\mathbf{f}_{k+1}^{(3)},
~\mathbf{P}_3=\mathbf{f}_{k+4}^{(1)}+\mathbf{f}_{k+1}^{(2)}+\mathbf{f}_{k+2}^{(3)},~\mathbf{P}_4=\mathbf{f}_{k+1}^{(1)}+\mathbf{f}_{k+2}^{(2)}+\mathbf{f}_{k+3}^{(3)}\\
\mathbf{P}_1=\mathbf{f}_{k+4}^{(1)}+\mathbf{f}_{k+3}^{(2)}+\mathbf{f}_{k+2}^{(3)},~\mathbf{P}_2=\mathbf{f}_{k+1}^{(1)}
+\mathbf{f}_{k+4}^{(2)}+\mathbf{f}_{k+3}^{(3)},
~\mathbf{P}_3=\mathbf{f}_{k+2}^{(1)}+\mathbf{f}_{k+1}^{(2)}+\mathbf{f}_{k+4}^{(3)},~\mathbf{P}_4=\mathbf{f}_{k+3}^{(1)}+\mathbf{f}_{k+2}^{(2)}+\mathbf{f}_{k+1}^{(3)}
\end{eqnarray*}
and
\begin{eqnarray*}
\mathbf{P}_1=\mathbf{f}_{k+2}^{(1)}+\mathbf{f}_{k+3}^{(1)}+\mathbf{f}_{k+4}^{(1)},~\mathbf{P}_2=\mathbf{f}_{k+1}^{(1)}
+\mathbf{f}_{k+3}^{(2)}+\mathbf{f}_{k+4}^{(2)},
~\mathbf{P}_3=\mathbf{f}_{k+1}^{(2)}+\mathbf{f}_{k+2}^{(2)}+\mathbf{f}_{k+4}^{(3)},~\mathbf{P}_4=\mathbf{f}_{k+1}^{(3)}+\mathbf{f}_{k+2}^{(3)}+\mathbf{f}_{k+3}^{(3)}
\end{eqnarray*}
where the fist two  are based on the injective functions \eqref{Eqn_MainD} and \eqref{Eqn_MinorD} respectively.

For these new three $4$-piggybaced $(k+4,k)$ MSR codes, the data amount of a node is $\alpha=4\alpha'$ and the repair bandwidth  of the parity nodes is $(k+12)\alpha/4$, which is asymptotically optimal with respect the bound $(k+3)\alpha/4$  given in \eqref{repair_bandwidth_MSR_bound}.

\end{Example}

\begin{Remark}
Recall that  the $(k+2,k)$ Hadamard MSR code is special for its optimal repair property of any node. In the following table, we give a comparison between the $(k+2,k)$ Hadamard MSR code, the $(k+2,k)$ Piggybacked Zigzag code and the $(k+2,k)$ Piggybacked Long MDS code. For simplicity, the same data amount $\alpha=2^m$ is adopted.

\begin{table}[htbp]
\begin{center}
\caption{Comparison of $(k+2,k)$ Hadamard MSR code to $(k+2,k)$ Piggybacked MSR codes}\label{Comparison_Hadamard_Piggyback-1}
\begin{tabular}{|c|c|c|c|}
\hline
& Hadamard & Piggybacked Zigzag & Piggybacked Long MDS \\
\hline
$\alpha$ & $2^m$ & $2^m$ & $2^m$ \\
\hline
$k$ & $m-1$ & $m$ & $3(m-1)$ \\
\hline
$\gamma_{\mathrm{system}}$ & $(k+1)\alpha/2$ & $(k+1)\alpha/2$ & $(k+1)\alpha/2$ \\
\hline
$\gamma_{\mathrm{parity}}$ & $(k+1)\alpha/2$ & $(k+2)\alpha/2$ & $(k+2)\alpha/2$ \\
\hline
\end{tabular}
\end{center}
\end{table}

From Table \ref{Comparison_Hadamard_Piggyback-1}, we can see that both of the two piggybacked codes, whose $\gamma_{\mathrm{system}}$ and $\gamma_{\mathrm{parity}}$ are $(k+1)\alpha/2$ and $(k+2)\alpha/2$ respectively, have optimal repair property of the systematic nodes and almost optimal repair property of the parity nodes, with respect to the theoretic bound in \eqref{repair_bandwidth_MSR_bound}. For a given $\alpha=2^m$, the piggybacked Zigzag code and the piggybacked Long MDS code respectively support $1$ and $2(m-1)$ more nodes than the Hadamard code. Hence, our new piggybacking method shows a clear advantage over the number of systematic nodes.
\end{Remark}

\section{Piggybacked $(k+r,k)$ MSR code with Balanced Download}\label{section_symmetric_piggyback}

According to the repair strategy in the last section, node repair of the $r$-piggybacked $(k+r,k)$ MSR code has the following characteristics:
\begin{enumerate}
\item [C1] To repair a systematic node, $\beta_1={\alpha' \over r}$ data should be downloaded from each surviving node;
\item [C2] To repair a parity node, $\beta_1={\alpha' \over r}$ and $\beta_2=\alpha'$ data should be downloaded from each systematic node and each surviving parity node respectively,
\end{enumerate}
where $\alpha'$ is the data amount of each node.
In the sense of load balance, the download is not balanced. In this section, by using the so-called layer technique presented  in \cite{tian,layer} , we give a construction of the piggybacked $(k+r,k)$ MSR code with balanced download.

\begin{Definition}\label{BIBD}(\cite{handbook})
An $(n,r,\lambda;e,b)$ balanced incomplete block design (BIBD) is a pair $(I_n,\mathcal{A})$ where $I_n$ is an $n$-set and $\mathcal{A}$
is a collection of $b$ $r$-subsets of $I_n$ (blocks) such that each element of $I_n$ is contained in exactly $e$ blocks and any $2$-subset of $I_n$ is contained in exactly $\lambda$ blocks.
\end{Definition}

For an $(n,r,\lambda;e,b)$-BIBD, the parameters $e,b$ can be determined by the other three as
\begin{equation}\label{lambda}
  e=\frac{\lambda(n-1)}{r-1},\,\,b=\frac{\lambda n(n-1)}{r(r-1)}.
\end{equation}
For simplicity, we use $(n,r,\lambda)$-BIBD to denote $(n,r,\lambda;e,b)$-BIBD.

\vspace{2mm}
Based on BIBDs, we are able to construct the piggybacked $(k+r,k)$ MSR code with balanced download as follows:
\begin{itemize}
\item [Step 1] Choose an $(n=k+r,r,\lambda)$-BIBD $(I_n,\mathcal{A})$, where $I_n=\{1,2,\cdots,n\}$ and $\mathcal{A}=\{A_i|1\le i\le b\}$;
\item [Step 2]  Generate $b$ instances of a $r$-piggybacked $(k+r,k)$ MSR code;
\item [Step 3]  For each $1\le i\le b$, distribute $r$ parity data and $k$ systematic data of the $i$th instance onto nodes in $A_i$ and $I_n\backslash A_i$ respectively, we then obtain a $(b\cdot r)$-piggybacked $(k+r,k)$ MSR code.
\end{itemize}

Below is an example of the new code.

\begin{Example}
Let $(I_{13},\mathcal{A})$ be an $(k+r=13,r=4,\lambda=1)$-BIBD defined by
\begin{equation*}
  A_i=\{(i-1)\%13+1,i\%13+1,(i+2)\%13+1,(i+8)\%13+1\},\,\,1\le i\le b=13
\end{equation*}
where $\%$ denotes the modulo operation.
A new $(13r)$-piggybacked $(k+r,k)$ MSR code with balanced download can be expressed by a matrix as
\begin{equation*}%\label{matrix}
\left(
\begin{array}{*{13}{c}}
 1 & 0 & 0 & 0 & 1 & 0 & 0 & 0 & 0 & 0 & 1 & 0 & 1  \\
 1 & 1 & 0 & 0 & 0 & 1 & 0 & 0 & 0 & 0 & 0 & 1 & 0  \\
 0 & 1 & 1 & 0 & 0 & 0 & 1 & 0 & 0 & 0 & 0 & 0 & 1  \\
 1 & 0 & 1 & 1 & 0 & 0 & 0 & 1 & 0 & 0 & 0 & 0 & 0  \\
 0 & 1 & 0 & 1 & 1 & 0 & 0 & 0 & 1 & 0 & 0 & 0 & 0  \\
 0 & 0 & 1 & 0 & 1 & 1 & 0 & 0 & 0 & 1 & 0 & 0 & 0  \\
 0 & 0 & 0 & 1 & 0 & 1 & 1 & 0 & 0 & 0 & 1 & 0 & 0  \\
 0 & 0 & 0 & 0 & 1 & 0 & 1 & 1 & 0 & 0 & 0 & 1 & 0  \\
 0 & 0 & 0 & 0 & 0 & 1 & 0 & 1 & 1 & 0 & 0 & 0 & 1  \\
 1 & 0 & 0 & 0 & 0 & 0 & 1 & 0 & 1 & 1 & 0 & 0 & 0  \\
 0 & 1 & 0 & 0 & 0 & 0 & 0 & 1 & 0 & 1 & 1 & 0 & 0  \\
 0 & 0 & 1 & 0 & 0 & 0 & 0 & 0 & 1 & 0 & 1 & 1 & 0  \\
 0 & 0 & 0 & 1 & 0 & 0 & 0 & 0 & 0 & 1 & 0 & 1 & 1
\end{array}
\right)
\end{equation*}
where the row denotes the node, the column denotes the instance of the $r$-piggybacked MSR code,  and $0,1$ denote systematic data and parity data respectively.
\end{Example}

Assume that the data amount of each node of the $r$-piggybacked $(k+r,k)$ MSR code is $\alpha'$. If node $i$ fails, following the repair strategy in the last section we repair the failed node of the new $(b\cdot r)$-piggybacked $(k+r,k)$ MSR code instance by instance. According to C1 and C2, we download data from node $j\ne i$ with instance $l$ ranging from $1$ to $b$ as follows:
\begin{enumerate}
\item [(1)] If node $i$ in  instance $l$ is a systematic node, ${\alpha' \over r}$ data is downloaded;
\item [(2)] If node $i$ in instance $l$ is a parity node,
\begin{enumerate}
\item [(2.1)] If node $j$ in instance $l$ is a systematic node, ${\alpha' \over r}$  data is downloaded;
\item [(2.2)] If node $j$ in instance $l$ is a parity node, $\alpha' $  data is downloaded.
\end{enumerate}
\end{enumerate}
Note that for the new code each node contains $e$ parity data and $b-e$ systematic data by Definition \ref{BIBD}.
Thus, the download
\begin{eqnarray*}
\beta&=&|\{l|i,j\in A_l,1\le l\le b\}|\cdot\alpha'+(e-|\{l|i,j\in A_l,1\le l\le b\}|)\cdot {\alpha' \over r}+(b-e){\alpha' \over r}\\
&=&|\{l|i,j\in A_l,1\le l\le b\}|\cdot\alpha'+(b-|\{l|i,j\in A_l,1\le l\le b\}|)\cdot {\alpha' \over r}\\
&=&\lambda\cdot\alpha'+(b-\lambda)\cdot {\alpha' \over r}\\
&=&{(b+(r-1)\lambda)\cdot \alpha \over b r}\\
&=&\left(\frac{(r-1)^2}{n(n-1)}+\frac{1}{r}\right)\alpha
\end{eqnarray*}
follows from Definition \ref{BIBD}, which implies $|\{t|i,j\in A_t,1\le t\le b\}|=\lambda$, and (\ref{lambda})  where $\alpha=b\alpha'$ is the data
amount of each node  in the new $(b\cdot r)$-piggybacked $(k+r,k)$ MSR code. This is to say,
the download $\beta$ is a constant independent of nodes $i$ and $j$ and hence is balanced.

\begin{Theorem}
The $(b\cdot r)$-piggybacked $(n=k+r,k)$ MSR code constructed above has the balanced download property that when repairing a failed node, the amount of data that each surviving node needs to transmit is $\beta=\left(\frac{(r-1)^2}{n(n-1)}+\frac{1}{r}\right)\alpha$, which is $\left(1+\frac{r(r-1)^2}{n(n-1)}\right)$ times of the optimal case.
\end{Theorem}

\section{Conclusion}\label{section_of_conclusion}
In this paper, we gave a systematic discussion on the design of piggybacking to achieve the minimal average repair bandwidth of parity nodes of MSR codes, and then presented a new piggybacking design which can give MSR codes with almost optimal repair bandwidth of parity nodes, while retaining the optimal repair bandwidth of systematic nodes. Moreover, a  construction of MSR codes with balanced download during node repair process based on the new piggybacking design was also proposed.

\end{document}